\documentclass[10pt]{article}
\usepackage{epsfig}
\usepackage{mathtools}
\usepackage{amsfonts}
\usepackage{amssymb}
\usepackage{amsmath}
\usepackage{shuffle} 
\usepackage{color}
\usepackage{slashed}
\usepackage{cite}
\usepackage{caption}
\usepackage{subcaption}

\usepackage{geometry} 
\usepackage{graphicx}
\graphicspath{{pictures/}}

\usepackage{multirow}           
\usepackage{array}              

\def\ep  {\varepsilon}
\def\E4  {\mathcal{E}_4}

\begin{document}

\newcommand{\Ef}[4]{\mathcal{E}_4\bigl(\begin{smallmatrix} #1 \\ #2 \end{smallmatrix};#3; \vec{#4}\bigl)}

\begin{center}

\vspace{3cm}

{\bf \Large Calculation of master integrals in terms of elliptic multiple polylogarithms} \vspace{1cm}

{\large M.A. Bezuglov$^{1,2}$}\vspace{0.5cm}

{\it $^1$Bogoliubov Laboratory of Theoretical Physics, Joint
    Institute for Nuclear Research, Dubna, Russia, \\
    $^2$Moscow Institute of Physics and Technology (State University), Dolgoprudny, Russia,}\vspace{1cm}
\end{center}

\begin{abstract}
In modern quantum field theory, one of the most important tasks is the calculation of loop integrals. Loop integrals appear when evaluating the Feynman diagrams with one or more loops by integrating over the internal momenta. Even though this problem has already been in place since the mid-twentieth century, we not only do not understand how to calculate all classes of these integrals beyond one loop, we do not even know in what class of functions the answer is expressed. To partially solve this problem, different variations of new functions called usually elliptic multiple polylogarithms have been introduced in the last decade. In this paper, we explore the possibilities and limitations of this class of functions. As a practical example, we chose the processes associated with the physics of heavy quarkonium at the two-loop level.

\end{abstract}

\begin{center}
\end{center}

\newpage

\tableofcontents{}\vspace{0.5cm}

\renewcommand{\theequation}{\thesection.\arabic{equation}}

\section{Introduction}
In modern quantum field theory, the main observable quantities are the scattering amplitudes that determine the probability of micro-level processes. The scattering amplitudes are usually considered within the framework of perturbation theory, i.e., the desired quantity is decomposed in a series according to the coupling constant, this coupling constant is regarded as small. Each $k$-th element of the perturbation series is represented as the sum of Feynman diagrams with $k$ loops. 
In order to solve them, it is necessary to calculate the integral over momenta flowing each of the $k$ loops - such integrals are called Feynman loop integrals. Previously, it was rarely required to calculate Feynman diagrams with two or more loops for complex processes. But in the past few decades, the accuracy of the measurements in particle physics has grown significantly. The commissioning of such machines as the Large Hadron Collider makes it necessary for the calculation of NNLO corrections for many measurable processes, for example, see \cite{gehrmann2014qcd, grazzini2016w}; therefore, it became necessary to develop methods for calculating Feynman integrals for two or more loops.

Each Feynman integral belongs to a specific family of integrals. We call family all integrals with the same structure of propagators but with arbitrary degrees of these propagators, including subgraphs. Elements of one family are not independent. There are so-called integration by parts(IBP) dependencies \cite{tkachov1981theorem,chetyrkin1981integration,laporta2000high} that establish a linear relationship between integrals of one family. These relations leads to the fact that any integral from this family can be represented as a linear combination of some limited basis of integrals\footnote{it is a basis in the full sense since it can be selected arbitrarily}, elements of this basis are called master integrals. Thus, to calculate integrals related to the scattering amplitude, one needs to determine the corresponding family of integrals, then introduce the basis of master integrals and calculate them. 

There are two main ways to calculate masterintegrals. The most modern method is to write a system of differential equations for a system of basis integrals\cite{KOTIKOV1991158,kotikov1991differential,kotikov1991differential2,remiddi1997differential,gehrmann2000differential, argeri2007feynman,henn2015lectures}. In this case, the master integrals basis is chosen so that the corresponding system of differential equations can be easily integrated.  The second method is a direct integration, which consists in introducing some parametric transformation, for example, Feynman, and then integrating over parameters. In this paper, we will use this second method.

\begin{figure}
\center{\includegraphics[width=0.8\textwidth]{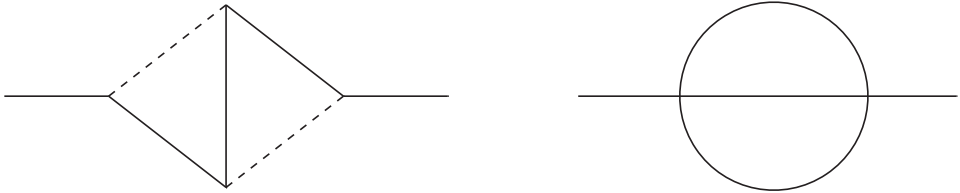}}
\caption{Kite integral on the left and sunset integral on the right. Dashed lines denote massless propagators and  thick lines represent massive propagators.}
\label{KiteAndSunset}
\end{figure}

Feynman integrals are usually expressed in terms of special functions. The most common is the so-called multiple polylogarithms(MPLs)\cite{goncharov2,goncharov3}, which are the natural generalization of ordinary logarithms. For MPLs, there are many functional dependencies, mainly because they form a Hopf algebra\cite{goncharov4}. These dependencies allow us to successfully use them to solve a large number of practical problems. Nevertheless, it is known for certain that not all Feynman integrals above one loop can be solved in terms of ordinary MPLs. One of the first examples was so-called kite integral(see the left part of Figure \ref{KiteAndSunset}) which appears in calculation of electron self-energy in QED at the two-loop level. This problem was first considered in\cite{sabry1962fourth}. Became known that the solution for an integral of the kite type should contain integrals of elliptic functions.  After this, similar problems often appeared in calculations related to the Standard Model and its extensions, the most simple example is the sunset integral with three massive lines(see the right part of Figure \ref{KiteAndSunset}) \cite{remiddi2014schouten,laporta2005analytic, bloch2015elliptic, remiddi2016differential, hidding2019all,
remiddi2017elliptic, kniehl2006two, adams2016walk, adams2014two, adams2015two, adams2016iterated, adams2013two,
broedel2018elliptic}. Thus, it becomes necessary to consider a certain generalization of ordinary MPLs. It is natural to call such generalization elliptic multiple polylogarithms(eMPLs) \cite{brown2011multiple}. There are many different ways to introduce  eMPLs, the most prominent is the form of iterated integrals \cite{brown2011multiple,hidding2019all,broedelPure,broedel2018elliptic2,broedel2018elliptic, broedel2019elliptic, adams2018feynman, adams1807class, ablinger2018iterated}. Elliptic multiple polylogarithms also can be considered as a generalization of polylogarithmic series\cite{ adams2016walk, adams2014two, adams2015two, adams2016iterated, adams2013two, adams2016kite}. In this paper, we will use the so-called pure eMPLs from the papers \cite{broedelPure, Broedel:2019hyg}. We believe that this particular class of functions gives the most convenient and compact results in solving practical problems. It is also important to note that eMPLs, as well as ordinary MPLs, satisfy the Hopf algebra \cite{broedelPure, broedelEllipticSymbols}, which potentially means that they can be successfully used in future practice.

The purpose of this paper is to use the methods from \cite{broedelPure, Broedel:2019hyg} to calculate some two-loop elliptic Feynman diagrams which arises when considering the physics of heavy quarkonium \cite{chen2017two, chen2018two} and to show that in the end, it’s necessary to introduce functions that are more general than eMPLs into consideration\footnote{Conclusions about the presence of more complex curves in particle physics were also made in \cite{huang2013genera, hauenstein2015global, bourjaily2018traintracks, brown2012k3, bloch2015feynman, georgoudis2015two, bloch2017local}.}.
The remainder of the paper is organized as follows. In section \ref{functions}, we give a brief overview of the main classes of functions that we will use in this paper. Next, in sections \ref{Section1}, we will examine in detail the example of a "triangle with one massless line and massive loop". This integral is related to the processes of CP-even heavy quarkonium productions and decays\cite{chen2017two, chen2018two}. In the next two sections \ref{I2I3} and \ref{I4}, we will also consider the integrals arising in the processes associated with heavy quarkonium production\cite{chen2017two, chen2018two}. Next, in section \ref{I5} we will consider an example of a "triangle with all massive lines and massive loop", In work \cite{hidding2019all}, integrals of this type were called next to linear reducible. Finally, in section \ref{Conclusions} we will discuss our results and give an outlook for the future.

\section{Class of functions}
\label{functions}
In this section, we shall review the main classes of functions that we will use in subsequent sections. As already mentioned, those functions are the (MPLs) and there elliptic extensions (eMPLs).

\subsection{Multiple polylogarithms}
There are many different notations for polylogarithms \cite{MPLsHopf1}, in this paper we will use so called Goncharov multiple polylogarithms \cite{goncharov2,goncharov3}. They can be defined recursively: 
\begin{equation}
\label{MPL_Def}
G(a_1,...,a_n;x)=\int\limits_0^x \frac{G(a_2,...,a_n;x')}{x'-a_1}dx', \qquad n>0,
\end{equation}
where $a_i,x \in \mathbb{C}$ , $n \in \mathbb{N}$- is called the weight and the recursion starts with $G(;x)=1$. The form of MPL $G(a_1,...,a_n;x)$ in which all $a_i$ are independent from $x$ is called canonical. 

This definition has one problem. If all $a_i$ are equal to zero then the integral in \ref{MPL_Def} becomes infinite. One can avoid this problem by introducing the special definition for this case 

\begin{equation}
G(\vec{0}_n;x)=\frac{\log^n x}{n!},
\end{equation}
where $\vec{0}_n$ denotes a sequence of n zeros.

MPLs can be connected to the "classical" polylogarithms by the relation
\begin{equation}
\text{Li}_n(x)=-G\left(\vec{0}_{n-1},\frac{1}{x};1\right)=\int\limits_0^x\frac{dx'}{x'}\text{Li}_{n-1}(x').
\end{equation}

The product of two MPLs ending by the same variable can be written as:
\begin{equation}
G(\vec{v},x)G(\vec{u},x)=\sum\limits_{\vec{c}=\vec{v}\shuffle \vec{u}}G(\vec{c},x),
\end{equation}
where $\shuffle$ denotes the shuffle product. 
Consider $u$ and $v$ to be some arbitrary words of length $n$ and $m$. The shuffle product $u \shuffle v$ is a sum of over $\frac{(m+n)!}{m!n!}$ possible permutations of letters of this words without changing the order of letters withing each word, For example:

\begin{equation}
ab \shuffle cd= abcd+acbd+cabd+acdb+cadb+cdab.
\end{equation} 

It is possible to give an alternative definition of the shuffle product based on the recurrence relations
\begin{equation}
u \shuffle \varnothing = \varnothing \shuffle u = u,
\end{equation} 
\begin{equation}
u\alpha \shuffle v\beta = (u \shuffle v\beta )\alpha +(u\alpha \shuffle v)\beta,
\end{equation} 
where $\alpha$ and $\beta$ are single elements, and $u$ and $v$ are arbitrary words. 

A very nice property of MPLs is that they form a closed space under primitives and derivatives. If $a_i$ are independent from $x$ and $R(x)$ is a rational function then the primitive of $R(x)G(\vec{a};x)$ can be expressed as a linear combination of some other MPLs  in which all coefficients and arguments are rational functions with respect to the variable $x$. Similarly, the derivative of $G(\vec{a}(x);f(x))$ also will be expressed as a linear combination of MPLs.  The latter follows directly from the differential equation
\begin{equation}
d G(a_1,...,a_n;x)=\sum\limits_{i=1}^nG(a_1,...,a_{i-1},a_{i+1},...,a_n;x)d \log \left(\frac{a_{i-1}-a_{i}}{a_{i+1}-a_i}\right),
\end{equation} 
with $a_i \ne a_{i \pm 1}$.

MPLs are a well-known class of functions and a full discussion of their properties goes far beyond the scope of this paper; a more detailed overview including MPLs Hopf algebra structure can be found in \cite{MPLsHopf1, Duhr1, Duhr2}.

\subsection{Elliptic multiple polylogarithms}
Before describing eMPLs, we give a brief description of the theory of elliptic functions. For convenience, hereinafter, we will use notations similar to the notations introduced in \cite{broedelPure}. All the theory concerning ordinary elliptic curves and simplest elliptic functions have been well known for more then a hundred years, so we will not give a detailed description of it, but restrict ourselves to recalling the main points. A more detailed description can be found, for example, in \cite{akhiezer1990elements, hurwitz1968function}.

We call the elliptic curve an equation $y^2=(x-a_1)(x-a_2)(x-a_3)(x-a_4)$, where $a_i$ are the complex numbers which are called the brunch points.

Next, we need to give definitions of some quantities that are associated with the elliptic curve and which we will need in the future.
First, we define the two periods of the elliptic curve as 

\begin{equation}
\omega_1=2c_4\int\limits_{a_2}^{a_3}\frac{dx}{y}=2\text{K}(\lambda), \qquad \omega_2=2c_4\int\limits_{a_1}^{a_2}\frac{dx}{y}=2i\text{K}(1-\lambda),
\end{equation}
where $\lambda=\frac{a_{14}a_{23}}{a_{13}a_{24}}$, $c_4=\sqrt{a_{13}a_{24}}$, $a_{ij}=a_i-a_j$ and $\text{K}$ denotes the complete elliptic integral of the first kind. The ratio of two periods  $\tau=\frac{\omega_2}{\omega_1}$ is called the module of the elliptic curve.  Note, that the ratio $\tau$ is a complex number i.e. $\text{Im}(\tau)\neq 0$.

Periods are important characteristics of an elliptic curve, but two different forms of the same elliptic curve with respect to the modular transformations may possess different periods. To avoid such complications it is necessary to introduce the concept of Weierstrass canonical form. 

Any elliptic curve can be transformed in to Weierstrass canonical form
\begin{equation}
\label{Weierstrass_canonical_form}
y^2=4x^3-g_2(\tau)x-g_3(\tau).
\end{equation}
One can obtain this form by using transformations which belongs to the modular group:
\begin{equation}
\label{ModTransform}
x\to \frac{ax-b}{cx-d}, \qquad y \to \frac{y}{(cx-d)^2}, \qquad ad-bc=1.
\end{equation}
The numbers $g_2(\tau)$ and $g_3(\tau)$ are known as the invariants of the elliptic curve.
 
From the invariants $g_2(\tau)$ and $g_3(\tau)$ we can compose a special function which is a modular form of weight zero and uniquely determines the isomorphism class of the elliptic curve. This function is called the $j$-invariant, and it is written as 
\begin{equation}
\label{jInvariant}
j(\tau)=1728\frac{g_2(\tau)^3}{g_2(\tau)^3-27g_3(\tau)^2}.
\end{equation}
If two curves have the same $j$-invariant then they are isomorphic to each other and can be reduced to the same form by  transformation \eqref{ModTransform}.

Equation \eqref{Weierstrass_canonical_form} describes a two-dimensional surface in four-dimensional space. In order to determine the type of this surface, one can use so-called Weierstrass $\wp$ function which is the simplest example of  elliptic function 
\begin{equation}
\wp(\mathit{z})=\frac{1}{\mathit{z}^2}+\sum_{(m,n)\ne (0,0)}\left(\frac{1}{(\mathit{z}+m\omega_1+n\omega_2)^2}-\frac{1}{(m\omega_1+n\omega_2)^2}\right),
\end{equation}
where the summation goes over all integers $m$ and $n$ excluding $(m,n)=(0,0)$ and $\omega_1$ and $\omega_2$ are the two periods such that $\text{Im}(\omega_2/\omega_1)\ne 0$. The Weierstrass $\wp$ function has one remarkable property, it is doubly periodic with respect to its periods $\wp(\mathit{z}+i\omega_1+j\omega_2)=\wp(\mathit{z})$ with $i,j \in \mathbb{Z}$. 

If we choose these periods as the periods of the elliptic curve, then it turns out that this elliptic curve can be parameterized in the following form
\begin{equation}
\wp'(\mathit{z})^2=4\wp(\mathit{z})^3-g_2(\tau)\wp(\mathit{z})-g_3(\tau).
\end{equation}
So it follows that the elliptic curve is isomorphic to a torus because  $\wp(\mathit{z})$ is doubly periodic.

After a brief description of the classical theory of elliptic curves, we proceed to describe the iterative integrals of elliptic curves. For eMPLs we use definition given in\cite{broedelPure}:
\begin{equation}
\label{eMPLdef}
\Ef{n1 & ... & n_k}{c_1 & ... & c_k}{x}{a}=\int\limits_0^x dx' \Psi_{n_1}(c_1,x',\vec{a})\Ef{n2 & ... & n_k}{c_2 & ... & c_k}{x'}{a}.
\end{equation}
The sum $\sum_i|n_i|$ is called the weight of the eMPL and the integration kernels $\Psi_{n_1}(c_1,x',\vec{a})$ are defined as
\begin{equation}
\label{Kernel0}
\Psi_{0}(0, x ,\vec{a})=\frac{c_4}{\omega_1 y}
\end{equation}
and
\begin{align}
\label{Kernel1}
\Psi_{1}(c, x ,\vec{a}) & =\frac{1}{x-c}, \qquad \Psi_{-1}(c, x ,\vec{a})=\frac{y(c)}{y(x-c)}+Z_4(c,\vec{a}), \qquad c\ne \infty,
\\
\Psi_{1}(\infty, x ,\vec{a}) & =-Z_4(x,\vec{a})\frac{c_4}{y}, \qquad  \Psi_{-1}(\infty, x ,\vec{a})=\frac{x}{y}-\frac{a_1+2c_4G_{*}(\vec{a})}{y}.
\end{align}
Note that from this definition directly follows that ordinary MPLs are a subset of eMPLs
\begin{equation}
\Ef{1 & ... & 1}{a_1 & ... & a_n}{x}{a}=G(a_1,...,a_n;x)
\end{equation}
and similar to MPLs, eMPLs defined by \eqref{eMPLdef} satisfy the shuffle algebra 
\begin{equation}
\label{E4Shuffle}
\mathcal{E}_4(\vec{V};x;a)\mathcal{E}_4(\vec{U};x;a)=\sum\limits_{\vec{C}=\vec{V}\shuffle \vec{U}}\mathcal{E}_4(\vec{C};x;a).
\end{equation} 

In order to define functions $Z_4(x,\vec{a})$ and $G_{*}(\vec{a})$ it is convenient to introduce the concept of
Eisenstein-Kronecker series
\begin{equation}
\label{EisensteinKronecker}
F(\mathit{z},\alpha,\tau)=\frac{1}{\alpha}\sum\limits_{n\ge 0}g^{(n)}(\mathit{z},\tau)\alpha^n=\frac{\theta_1'(0,\tau)\theta_1(\mathit{z}+\alpha,\tau)}{\theta_1(\mathit{z},\tau)\theta_1(\alpha,\tau)},
\end{equation}
where $\theta_1$ is the odd Jacobi theta function.

The functions $g^{(n)}(\mathit{z},\tau)$ possess a certain parity
\begin{equation}
\label{EisensteinKroneckerParity}
g^{(n)}(-\mathit{z},\tau)=(-1)^ng^{(n)}(\mathit{z},\tau),
\end{equation}
and with respect to translations by 1 and $\tau$, they behave as
\begin{equation}
\label{EisensteinKroneckerTranslations}
g^{(n)}(\mathit{z}+1,\tau)=g^{(n)}(\mathit{z},\tau), \qquad g^{(n)}(\mathit{z}+\tau,\tau)=\sum\limits_{k=0}^n\frac{(-2\pi i)^k}{k!}g^{(n-k)}(\mathit{z},\tau).
\end{equation}

Using generating series \eqref{EisensteinKronecker}, functions $Z_4(x,\vec{a})$ and $G_{*}(\vec{a})$ are defined as\cite{broedelPure}:
\begin{equation}
\label{Z4Eisenstein}
Z_4(x,\vec{a})=-\frac{1}{\omega_1}\left(g^{(1)}(\mathit{z}_{x}-\mathit{z}_{*},\tau)+g^{(1)}(\mathit{z}_{x}+\mathit{z}_{*},\tau)\right),
\end{equation}
and
\begin{equation}
\label{Gstar}
G_{*}(\vec{a})=\frac{g^{(1)}(\mathit{z}_{*},\tau)}{\omega_1},
\end{equation}
where $\mathit{z}_{x}$ is the image of point $x$ on a torus and $\mathit{z}_{*}$ is the image of point $x=-\infty$
\begin{equation}
\label{Torus_coordinates}
\mathit{z}_{x}=\frac{c_4}{\omega_1}\int\limits_{a_1}^x\frac{dx}{y},\qquad \mathit{z}_{*}=\frac{c_4}{\omega_1}\int\limits_{a_1}^{-\infty}\frac{dx}{y}.
\end{equation}
Such image is the inverse of the elliptic function that parametrizes the elliptic curve $y$ and called the Abel's map.

Now we will move on to the practical use of eMPLs defined in \eqref{eMPLdef} to solve specific problems.

\section{Triangle with one massless line and massive loop}
\label{Section1}
\begin{figure}[h]
\center{\includegraphics[width=0.5\textwidth]{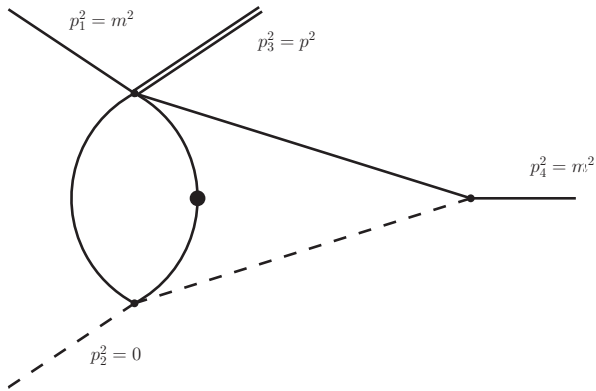}}
\caption{$I_1$ diagram. Dashed lines denote massless propagators and on-shell massless external particles; thick lines represent massive propagators and on-shell massive external particles; double line denotes off-shell external particle. A dot on a line means that the corresponding propagator is in the power two.}
\label{E7Diagram}
\end{figure}

As the first example, we will choose the integral described by the graph $I_1$ shown on Figure \ref{E7Diagram}. This graph, in particular, arises in the processes of production and decay of heavy quarkonium \cite{chen2017two, chen2018two}.
The corresponding Feynman integral reads

\begin{equation}
\label{I1}
I_1=e^{2\gamma_E\ep}(\mu^{2\ep})\int\frac{d^dk_1d^dk_2}{(i \pi^{d/2})^2}\frac{1}{(k_1^2-m^2)^2((k_1-k_2)^2-m^2)((p_1-p_3+k_2)^2-m^2)(p_2-k_2)^2},
\end{equation}
where $\gamma_E=-\Gamma'(1)$ is the Euler-Mascheroni constant and $d=4-2\ep$. We will carry out all subsequent calculations in the Euclidean region and put $t=-(p_1-p_3)^2>0$, the rest of the kinematics can be clearly understood from Figure \ref{E7Diagram}.

We can use the Feynman parametrization and rewrite the integral \eqref{I1} in a more convenient form

\begin{equation}
I_1=e^{2\gamma_E\ep}(\mu^{2\ep})\Gamma(1+2\ep)\int\limits_{\Delta}\left(\prod\limits_{i=1}^4dx_i\right) x_1 \frac{U^{5-\frac{3d}{2}}}{F^{5-d}},
\end{equation}
where $U$ and $F$ are the first and second Symanzik polynomials 
\begin{equation}
U=x_2x_3+x_2x_4+x_1x_2+x_1x_3+x_1x_4,
\end{equation}
\begin{equation}
\label{FforI1}
F=-tx_1x_2x_3-m^2\left((x_1+x_2+x_3)(x_2x_3+x_1x_2+x_1x_3)+x_4(x_1+x_2)^2\right)
\end{equation}
and the integration domain is $\Delta \in \left\{\vec{x}~|~x_i>0,~ \sum_{i=1}^4x_i=1 \right\}$. 

We may use the Cheng-Wu theorem to factor out one variable from the integral, this choice should be made in such way that the resulting integral can be integrated as simply as possible, in our case, the best choice would be the variable $x_1$:
\begin{equation}
\label{I_1UF}
I_1=e^{2\gamma_E\ep}\Gamma(1+2\ep)(\mu^{2\ep})\int\limits_{0}^{\infty}\prod\limits_{i=1}^4dx_i x_1 \frac{U^{5-\frac{3d}{2}}}{F^{5-d}}\delta(1-x_1).
\end{equation}

Since the integral $I_1$ is free from ultraviolet divergences in four dimensions, the expression \eqref{I_1UF} can be expanded in to a series in $\ep$
\begin{equation}
I_1=\left(-\frac{\mu^2}{m^2}\right)^{2\ep}\left[I_1^{(0)}+\ep I_1^{(1)} + \mathcal{O}(\ep^2)\right],
\end{equation}
where
\begin{equation}
I_1^{(0)}=\int\limits_0^{\infty}\frac{x_1 dx_1dx_2dx_3dx_4}{U F}\delta(1-x_1)=\int\limits_0^{\infty}\frac{dx_2dx_3dx_4}{U F |_{x_1=1}}.
\end{equation}
We'll start with analytical calculation of this integral

First of all, we see that the Symanzik polynomials are quadratic in the variables $x_1$, $x_2$ and $x_3$ and linear in the variable $x_4$, we have already excluded the variable $x_1$ using the Cheng-Wu theorem; therefore, the first integration must be carried out with respect to the variable $x_4$. 
We use the definition \eqref{MPL_Def} to integrate over the $x_4$ variable and we can immediately compute the primitive with respect to $x_4$

\begin{equation}
\label{primitive1E7}
\int\frac{dx_4}{U F |_{x_1=1}}=\frac{G\left(-\frac{tx_2x_3+m^2(1+x_2+x_3)(x_2+x_3+x_2x_3)}{m^2(1+x_2)^2};x_4\right)-G\left(-1+\frac{1}{1+x_2}-x_3;x_4\right)}{(1+x_2)x_3(t x_2+m^2(x_2+x_3+x_2x_3))}.
\end{equation}

Now we need to substitute the integration limits. The limit $x_4=0$ is trivial, so we focus on the limit $x_4=\infty$. In order to take this limit, we will perform the change of variable $x_4=\frac{1}{\epsilon}$ and focus on the limit $\epsilon \rightarrow 0$.  After that, we can use the fact that all multiple polylogarithms of weight one are just ordinary logarithms $G(a;b)=\log\left(1-\frac{b}{a}\right),~ a\ne 0$ and $G(0;b)=\log(b)$. Using elementary transformations we find 

\begin{equation}
\label{weight1Limit}
G\left(a;\frac{1}{\epsilon}\right)=i\pi-G(0;a)-G(0;0),\qquad \epsilon \rightarrow 0.
\end{equation}
here $G(0;0)$ is a logarithmic singularity $\log(\epsilon)$ which should cancel out in the final answer. Using this equation and the primitive \eqref{primitive1E7} we find
\begin{equation}
I_1^{(0)}=\int\limits_0^{\infty}dx_3dx_2\frac{G\left(0;-1+\frac{1}{1+x_2}-x_3\right)-G\left(0;-\frac{tx_2x_3+m^2(1+x_2+x_3)(x_2+x_3+x_2x_3)}{m^2(1+x_2)^2}\right)}{(1+x_2)x_3(t x_2+m^2(x_2+x_3+x_2x_3))}.
\end{equation}

The integrand is a linear combination of MPLs in which all coefficients and arguments are rational functions with respect to the variables $x_2$ and $x_3$, therefore, the result of the next integration will also lie in the class of MPLs.

We carry out the following integration with respect to the variable $x_3$. At this stage, we will also make the change of other variable $x_2=\frac{x}{1-x}$  so that the last integration domain will be in the range from 0 to 1. The transformation of this kind is called the M\"obius transformation and it's Jacobian is $J=\frac{1}{(1-x)^2}$. In order to carry out the last but one integration, we again use the properties of multi-polylogarithms of weight one to transform them to the canonical form $G(f;x_3)$ where $f$ does not depend on $x_3$ and we get

\begin{equation}
\label{e7_2int}
I_1^{(0)}=\int\limits_0^{\infty}dx_3dx_2\frac{G(-x;x_3)-G\left(\Upsilon(y);x_3\right)-G\left(\Upsilon(-y);x_3\right)}{(1+x_2)x_3(t x_2+m^2(x_2+x_3+x_2x_3))},
\end{equation}
where
\begin{equation}
\label{upsilon}
\Upsilon(y)=\frac{m^2(1+x-x^2)+tx(1-x)+(m^2+t)y}{2m^2(x-1)},
\end{equation}
and $y$ is the elliptic curve 
\begin{equation}
y^2=(x-a_1)(x-a_2)(x-a_3)(x-a_4),
\end{equation}
with the branch points
\begin{equation}
\label{sunsetBP}
a_1=\frac{1}{2}\left(1-\sqrt{1-\frac{4m^2}{(m+i\sqrt{t})^2}}\right),~~~a_2=a_1^*,~~~a_3=\frac{1}{2}\left(1+\sqrt{1-\frac{4m^2}{(m+i\sqrt{t})^2}}\right),~~~a_4=a_3^*.
\end{equation}
This is recognized immediately as the elliptic curve corresponding to the sunset graph \cite{broedelPure,broedel2018elliptic}. 

In equation \eqref{e7_2int} all MPLs are in canonical form so it can be further integrated and  we compute the primitive with respect to the variable $x_3$ of the integrand \eqref{e7_2int}
\begin{multline}
\frac{1}{(t+m^2)x}\left(G(0,-x;x_3)-G(0,\Upsilon(y);x_3)-G(0,\Upsilon(-y);x_3)-G\left(-\frac{(m^2+t)x}{m^2},-x;x_3\right)+\right.
\\
\left.+G\left(-\frac{(m^2+t)x}{m^2},\Upsilon(y);x_3\right)+G\left(-\frac{(m^2+t)x}{m^2},\Upsilon(-y);x_3\right)\right).
\end{multline}

Now we need to substitute the integration limit $x_3=\infty$ as we did it in equation \eqref{primitive1E7}, but now all MPLs have weight two, which somewhat complicates the matter. In order to substitute this limit, we use methods presented in article \cite{anastasiou2013soft}, based on the use of Hopf algebra for MPLs, this method allows us to reduce MPLs in to the canonical form, using it we find the following formula\footnote{This formula can also be obtained using special packages for MPLs, such as \cite{duhr2019polylogtools}.}
\begin{multline}
\label{weight2Limit}
G\left( a,b;\frac{1}{\epsilon}\right)=-\frac{\pi^2}{3}-i\pi(G(0;0)+G(0;b))+G(0;0)G(0;b)+\\+(G(0;b)-G(0;a))G(b;a)+G(0,0;0)+G(0,0;b)+G(0,b;a),
\end{multline}
where $G(0,0;0)=1/2G(0;0)^2$ is a logarithmic singularity.

And we find that
\begin{multline}
\label{e7_lastint}
I_1^{(0)}=\int\limits_0^{1}\frac{dx}{(m^2+t)x}\left(-G(0;-x)G\left(x;\frac{(m^2+t)x}{m^2}\right)+G(0;\Upsilon(y))
G\left(\Upsilon(y);-\frac{(m^2+t)x}{m^2}\right)+
\right.\\
+G(0;\Upsilon(-y))G\left(\Upsilon(-y);-\frac{(m^2+t)x}{m^2}\right)+G\left(0;-\frac{(m^2+t)x}{m^2}\right)\times
\\
\times\left(
G\left(x;\frac{(m^2+t)x}{m^2}\right)-G\left(\Upsilon(y);-\frac{(m^2+t)x}{m^2}\right)-G\left(\Upsilon(-y);-\frac{(m^2+t)x}{m^2}\right)\right)-
\\
-\left. G\left(0,x,\frac{(m^2+t)x}{m^2}\right)+G\left(0,\Upsilon(y),-\frac{(m^2+t)x}{m^2}\right)+G\left(0,\Upsilon(-y),-\frac{(m^2+t)x}{m^2}\right)\right).
\end{multline}

Note that all the singularities in this expression have been reduced and this expression is automatically symmetrical with respect to the rearrangements $y \leftrightarrow -y$. This is because the root $y$ originates from the second degree polynomial in $x_3$ from the equation \eqref{primitive1E7}:
\begin{multline}
\frac{tx_2x_3+m^2(1+x_2+x_3)(x_2+x_3+x_2x_3)}{m^2(1+x_2)^2}=\frac{t}{m^2}(x-1)xx_3+(x+x_3)\left((x-1)x_3-1\right)=
\\
=(x-1)\left(x_3-\Upsilon(y)\right)\left(x_3-\Upsilon(-y)\right),
\end{multline}
which preserves this symmetry.

Expression \eqref{e7_lastint} is an integral of an MPLs combination, however, it cannot be integrated further in the same class of functions because the arguments of these MPLs are not rational functions with respect to the integration variable. In order to move forward, it is necessary to rewrite each MPL in the expression \eqref{e7_lastint} as a linear combination of eMPLs. The corresponding recurrence algorithm was described in \cite{Broedel:2019hyg} and\cite{broedelII}. Its brief description is as follows, first, we take the full derivative of the selected MPL of weight $n$ with respect to the variable $x$, after which we get a linear combination of MPLs of weight $n-1$ and where all coefficients are rational functions with respect to the variable $x$ and the elliptic curve $y$. Since the algorithm is based on recursion, we can assume that all MPLs of weight $n-1$ are already rewritten as linear combinations of eMPLs, it is obvious that recursion begins with $n = 1$. Further, we can integrate the obtained combination of eMPLs using definition \eqref{eMPLdef}. In the end, we need to fix the integration constant. For this, we can compare the values of the initial expression and the final expression at some point, usually, as such a point, it is most convenient to choose $ x = 0$\footnote{For the point $x = 0$, this algorithm can be described by a slightly modified Newton-Leibniz formula $F_{MPL}(x)=\left.\int\limits_0^x\frac{dF_{MPL}(x)}{dx}dx\right|_{0}^x + F_{MPL}(0)$ where we need the additional limits $|_{0}^x$ In order to correctly subtract all divergences}.

Consider a simple example $G(0;\Upsilon(y))$, the full derivative reads

\begin{equation}
\frac{dG(0;\Upsilon(y))}{dx}=-\frac{1}{2(x-1)x}+\frac{1}{2y}-\frac{x}{y}+\frac{m^2(1-2x)}{2(m^2+t)y(x-1)x}.
\end{equation}

Now we can integrate this expression with the help of \eqref{eMPLdef} and use the boundary value $\left.G(0;\Upsilon(y))\right|_{x=0}=i\pi$, we find
\begin{equation}
\begin{split}
G(0;\Upsilon(y))=i\pi-\frac{1}{2}\Ef{-1}{0}{x}{a}-\frac{1}{2}\Ef{-1}{1}{x}{a}-\Ef{-1}{\infty}{x}{a}+\frac{1}{2}\Ef{1}{0}{x}{a}-\frac{1}{2}\Ef{1}{1}{x}{a}+\\+\frac{\omega_1}{2}\Ef{0}{0}{x}{a}\left\{\frac{1-2a_1}{c_4}-4G_{*}(\vec{a})+Z_4(0,\vec{a})+Z(1,\vec{a})\right\}.
\end{split}
\end{equation}
The last expression can be somewhat simplified with the help of the explicit forms of $G_{*}(\vec{a})$ and $Z_4(0,\vec{a}) + Z_4(1,\vec{a})$ for sunset topology which were obtained in \cite{broedelPure}. 
\begin{align}
G_{*}(\vec{a}) & =\frac{i\pi}{2\omega_1}-\frac{2 a_1-1}{4c_4},
\\
Z_4(0,\vec{a}) & + Z_4(1,\vec{a})=\frac{2\pi i}{\omega_1}.
\end{align}
Note, that in the reference \cite{broedelPure}. the last formula was obtained with the help of numerical analysis, in Appendix A we will show how this formula can be obtained analytically and for more general case.

And we find
\begin{equation}
G(0;\Upsilon(y))=i\pi-\frac{1}{2}\Ef{-1}{0}{x}{a}-\frac{1}{2}\Ef{-1}{1}{x}{a}-\Ef{-1}{\infty}{x}{a}+\frac{1}{2}\Ef{1}{0}{x}{a}-\frac{1}{2}\Ef{1}{1}{x}{a}.
\end{equation}
Applying this method to all $G$ functions in formula \eqref{e7_lastint} and using the shuffle algebra \eqref{E4Shuffle}  we obtain the following result
\begin{multline}
I_1^{(0)}=\int\limits_0^{1}\frac{-2dx}{(m^2+t)x}\left(
\Ef{-1 & -1}{\infty & 0}{x}{a}+\Ef{-1 & -1}{\infty & 1}{x}{a}+2\Ef{-1 & -1}{\infty & \infty}{x}{a}\right.+
\\
\left.
+i\pi(\Ef{0 & -1}{0 & 0}{x}{a}+\Ef{0 & -1}{0 & 1}{x}{a}+2\Ef{0 & -1}{0 & \infty}{x}{a})
\right).
\end{multline}

The final integration is trivial and we obtain the final answer

\begin{multline}
\label{I1FinalResult}
I_1^{(0)}=-\frac{2}{(m^2+t)}\left(
\Ef{1 & -1 & -1}{0 & \infty & 0}{x}{a}+\Ef{1 & -1 & -1}{0 & \infty & 1}{x}{a}+2\Ef{1 & -1 & -1}{0 & \infty & \infty}{x}{a}\right.+
\\
+\left.
i\pi(\Ef{1 & 0 & -1}{0 & 0 & 0}{x}{a}+\Ef{1 & 0 & -1}{0 & 0 & 1}{x}{a}+2\Ef{1 & 0 & -1}{0 & 0 & \infty}{x}{a})
\right).
\end{multline}

Similarly, we can obtain expressions for higher $\ep$ corrections, for example, correction $I_1^{(1)}$ is expressed by the following Feynman integral
\begin{equation}
I_1^{(1)}=\int\limits_0^{\infty}dx_1dx_2dx_3dx_4\frac{x_1\delta(1-x_1)\left(3 \log U-2 \log F  \right)}{U F}.
\end{equation}
This integral can be calculated using exactly the same methods as we used for $I_1^{(0)}$. The final result is very lengthy and can be found in supplementary notebook file.

Both of these results were verified numerically by comparison with direct numerical calculation of the parametric integrals. In this work, we will check all our results in a similar way, therefore we will not mention this furthermore.

\section{Triangles with different external lines}
\label{I2I3}
\begin{figure}[h]
\begin{minipage}[h]{0.5\linewidth}         
\center{\includegraphics[width=0.8\textwidth]{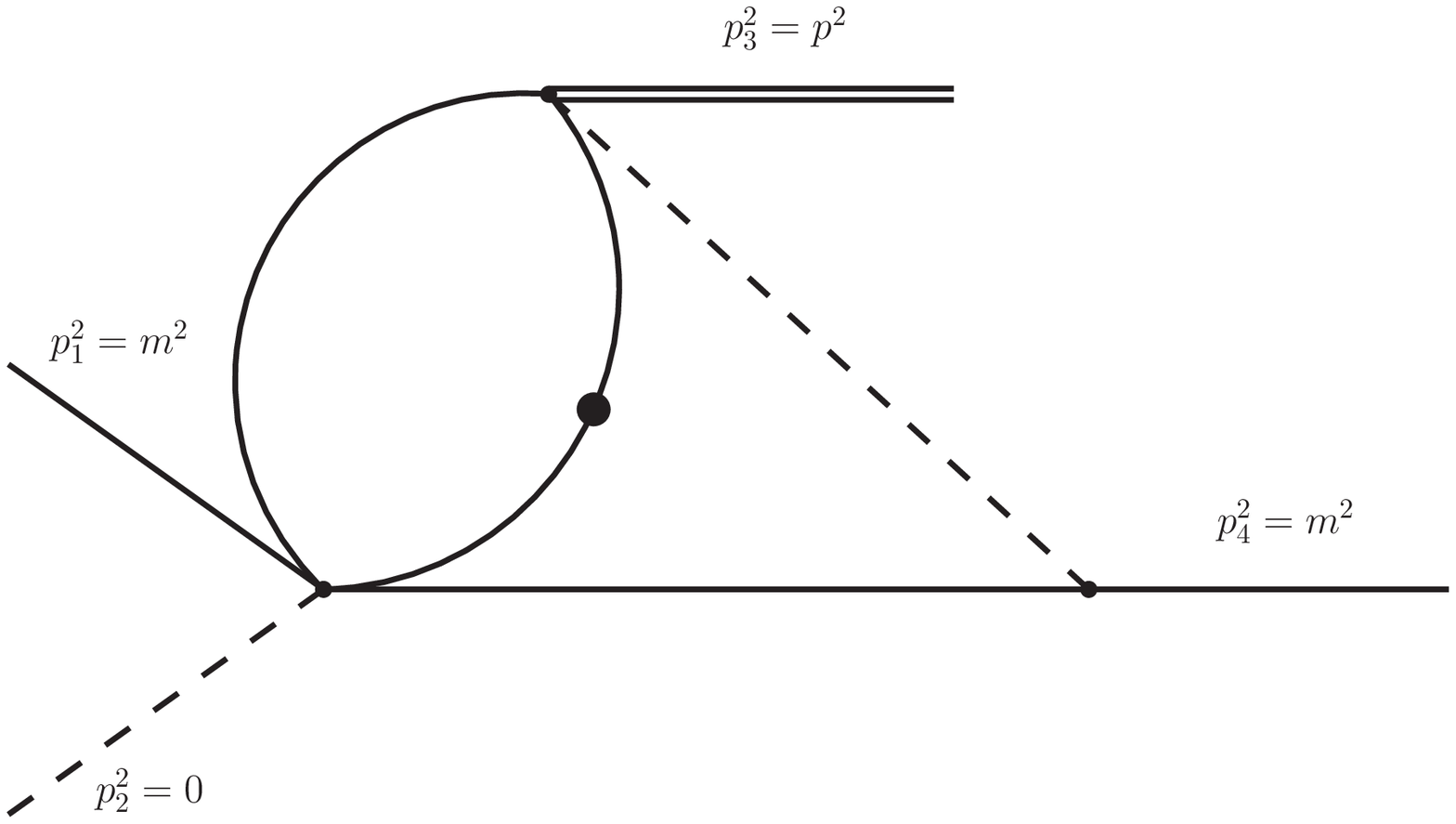}}
\end{minipage}
\hfill                             
\begin{minipage}[h]{0.5\linewidth}
\center{\includegraphics[width=0.8\textwidth]{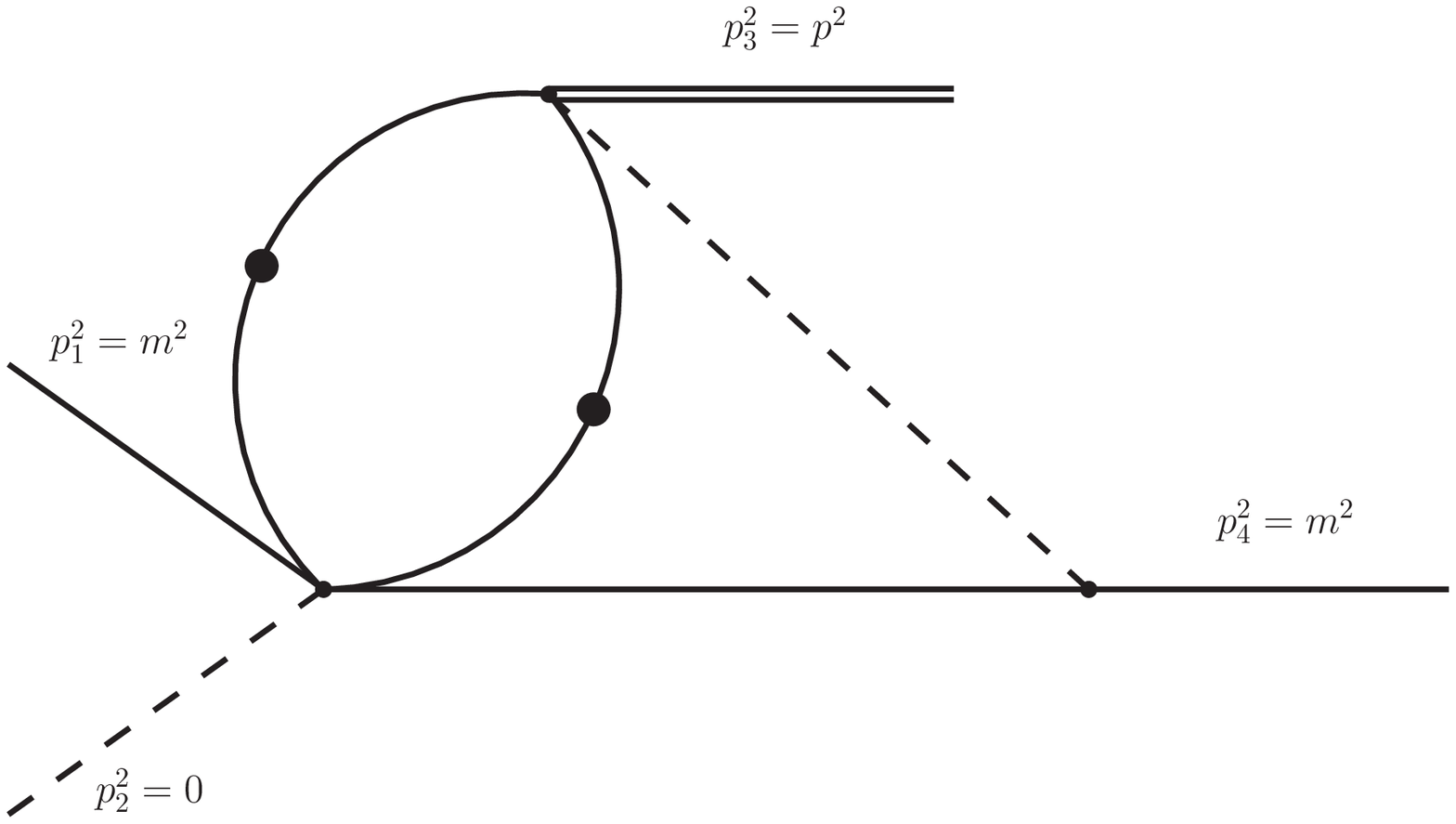}}
\end{minipage}
\caption{$I_2$ diagram on the left and $I_3$ diagram on the right. Dashed lines denote massless propagators and on-shell massless external particles; thick lines represent massive propagators and on-shell massive external particles; double line denotes off-shell external particle.}
\label{e8e9}
\end{figure}

As a second example, we consider two graphs shown in Figure \ref{e8e9}. These diagrams are very similar to the diagram from Figure \ref{E7Diagram} but have a different structure of external lines. The integral representation of $I_2$ is
\begin{equation}
I_2=e^{2\gamma_E\ep}(\mu^{2\ep})\int\frac{d^dk_1d^dk_2}{(i \pi^{d/2})^2}\frac{1}{(k_1^2-m^2)^2((k_1-k_2)^2-m^2)((p_1+p_2+k_2)^2-m^2)(p_3-k_2)^2}.
\end{equation}
The Symanzik polynomials are
\begin{equation}
\label{UforI2}
U=x_2x_3+x_2x_4+x_1x_2+x_1x_3+x_1x_4,
\end{equation}
\begin{equation}
\label{FforI2}
F=x_1x_2(p^2x_4-sx_3)-m^2\left((x_1+x_2+x_3)(x_2x_3+x_1x_2+x_1x_3)+x_4(x_1+x_2)^2\right),
\end{equation}
where $s=-(p_1+p_2)^2>0$ lies in the Euclidean region.

As before, we carry out the first integration with respect to the variable $x_4$ and substitute the integration limits using the formula \eqref{weight1Limit}, we find for $\ep^0$ correction
\begin{equation}
I_2^{(0)}=\int\limits_0^{\infty}dx_3\int\limits_0^1dx \frac{G\left(0;-x-x_3\right)-G\left(0;\frac{s(x-1)x x_3+m^2(x+x_3)(-1+(x-1)x_3)}{m^2+p^2(x-1)x}\right)}{p^2x(x+x_3)+x_3(sx+m^2(x+x_3))},
\end{equation}
here we have already replaced the variable $x_2=\frac{x}{1-x}$.

Further, we reduce all MPLs to canonical form and obtain
\begin{equation}
I_2^{(0)}=\int\limits_0^{\infty}dx_3\int\limits_0^1dx \frac{G(-x;x_3)-G\left(0;\frac{m^2}{m^2+p^2(x-1)x}\right)-G\left(\Upsilon(y);x_3\right)-G\left(\Upsilon(-y);x_3\right)}{p^2x(x+x_3)+x_3(sx+m^2(x+x_3))},
\end{equation}
where $\Upsilon(y)$ is defined by equation \eqref{upsilon} and $y^2=(x-a_1)(x-a_2)(x-a_3)(x-a_4)$,
with the branch points:
\begin{equation}
\label{brp2}
a_1=\frac{1}{2}\left(1-\sqrt{1-\frac{4m^2}{(m+i\sqrt{s})^2}}\right),~~~a_2=a_1^*,~~~a_3=\frac{1}{2}\left(1+\sqrt{1-\frac{4m^2}{(m+i\sqrt{s})^2}}\right),~~~a_4=a_3^*.
\end{equation}
We see that these branch points differ from \eqref{sunsetBP} only by replacement of $t$ by $s$. Therefore, we will denote the brunch points \eqref{brp2} with the same letter, which should not lead to confusion.

Using definition \eqref{MPL_Def}, we integrate over the variable $x_3$ and substitute the integration limits according to the formula \eqref{weight2Limit}

\begin{multline}
\label{I2LastInt}
I_2^{(0)}=\int\limits_0^1\frac{dx}{\eta x}\Bigg\{ -G \left(0;-\frac{m^2 x}{m^2+p^2(x-1)x}\right)G(0;\Psi_{\pm})+G(0;-x)\left(G(0;\Psi_{\pm})-G(-x;\Psi_{\pm})\right)+
\\
+G(0;\Upsilon(y))G(\Upsilon(y);\Psi_{\pm})+G(0;\Upsilon(-y))G(\Upsilon(-y);\Psi_{\pm})+G(-x,0;\Psi_{\pm})-
\\
-G(\Upsilon(y),0;\Psi_{\pm})-G(\Upsilon(-y),0;\Psi_{\pm})\Bigg\},
\end{multline}
where we introduce the shorten notations 
\begin{equation}
G\left(\vec{a};\Psi_{\pm}\right)=G\left(\vec{a};\Psi(\eta)\right)-G\left(\vec{a};\Psi(-\eta)\right)
\end{equation}
and
\begin{equation}
\Psi(\eta)=-\frac{m^2+p^2+s+\eta}{2m^2}, \qquad \eta=\sqrt{\left((m-p)^2+s\right)\left((m+p)^2+s\right)}.
\end{equation}
Note that \eqref{I2LastInt} is automatically symmetric with respect to permutations $y \leftrightarrow -y$ and $\eta \leftrightarrow -\eta $.

Rewriting all MPLs in expression \eqref{I2LastInt} through a linear combination of eMPLs and performing the last trivial integration, we find

\begin{equation}
I_2^{(0)}=-\frac{1}{\eta}\left[A_+ + A_-  - \omega_1\left(Z_4\left(\kappa_{+},\vec{a}\right)+Z_4\left(\kappa_{-},\vec{a}\right)\right) A_0\right],
\end{equation}
where we have introduced the notations
\begin{equation}
A_{\pm}=
\Ef{1 & -1 & -1}{0 & \kappa_{\pm} & 0}{1}{a}+\Ef{1 & -1 & -1}{0 & \kappa_{\pm} & 1}{1}{a}+2\Ef{1 & -1 & -1}{0 & \kappa_{\pm} & \infty}{1}{a},
\end{equation}
\begin{equation}
A_{0}=
\Ef{1 & 0 & -1}{0 & 0 & 0}{1}{a}+\Ef{1 & 0 & -1}{0 & 0 & 1}{1}{a}+2\Ef{1 & 0 & -1}{0 & 0 & \infty}{1}{a},
\end{equation}
\begin{equation}
\kappa_{\pm}=\frac{1}{2}\pm\frac{1}{2}\sqrt{1-\frac{4m^2}{p^2}}.
\end{equation}
This result can be further simplified, using results from Appendix A we find
\begin{equation}
\label{z4ForI2}
Z_4\left(\kappa_{+},\vec{a}\right)+Z_4\left(\kappa_{-},\vec{a}\right)=\frac{2\pi i}{\omega_1}
\end{equation}
and the final answer reads
\begin{equation}
\label{I2FinalResult}
I_2^{(0)}=-\frac{1}{\eta}\left[A_+ + A_-  -2\pi i  A_0\right].
\end{equation}

Next, we will consider diagram on the right part of Figure \ref{e8e9}. The corresponding Feynman integral reads
\begin{equation}
I_3=e^{2\gamma_E\ep}(\mu^{2\ep})\int\frac{d^dk_1d^dk_2}{(i \pi^{d/2})^2}\frac{1}{(k_1^2-m^2)^2((k_1-k_2)^2-m^2)^2((p_1+p_2+k_2)^2-m^2)(p_3-k_2)^2}.
\end{equation}

For the $\ep^0$ contribution one can find the following Feynman parametric representation
\begin{equation}
I_3^{(0)}=\int\limits_0^{\infty}\frac{x_1 x_2 dx_1dx_2dx_3dx_4}{F^2}\delta(1-x_1),
\end{equation}
where $U$ and $F$ are the Symanzik polynomials from \eqref{UforI2} and \eqref{FforI2}. Taking trivial integrals over the variables $x_4$ and $x_3$, we obtain the following formula
\begin{equation}
\label{I3LastInt}
I_3^{(0)}=\int_0^1\frac{x(x-1)\left(G\left(0;\Upsilon(y)\right)-G\left(0;\Upsilon(-y)\right)\right)}{(m^2+s)(m^2+p^2x(x-1))y}dx,
\end{equation}
where we already perform transformation $x_2=\frac{x}{1-x}$. Differentiating with respect to the variable $x$ and integrating back we find
\begin{equation}
G\left(0;\Upsilon(y)\right)-G\left(0;\Upsilon(-y)\right)=G\left(0;\frac{\Upsilon(y)}{\Upsilon(-y)}\right)=-\Ef{-1}{0}{x}{a}-\Ef{-1}{1}{x}{a}-2\Ef{-1}{\infty}{x}{a}.
\end{equation}
Inserting this equation in to expression \eqref{I3LastInt} and carrying out the last simple integration we find 
\begin{equation}
I_3^{(0)}=\frac{S_+-S_-}{p^2(\kappa_+-\kappa_-)\eta}-\left(\frac{\omega_1}{p^2(m^2+s)c_4(\vec{a})}+\frac{\omega_1(Z_4(\kappa_+,a)-Z_4(\kappa_-,a))}{p^2(\kappa_+-\kappa_-)\eta}\right)S_0,
\end{equation}
where we introduced the following definitions
\begin{equation}
S_{\pm}=\Ef{ -1 & -1}{\kappa_{\pm} & 0}{1}{a}+\Ef{-1 & -1}{\kappa_{\pm} & 1}{1}{a}+2\Ef{-1 & -1}{\kappa_{\pm} & \infty}{1}{a}
\end{equation}
and
\begin{equation}
S_{0}=\Ef{ 0 & -1}{0 & 0}{1}{a}+\Ef{0 & -1}{0 & 1}{1}{a}+2\Ef{0 & -1}{0 & \infty}{1}{a}.
\end{equation}

We also calculated the $\ep$ corrections $I_2^{(1)}$ and $I_3^{(1)}$ for the integrals in Figure \ref{I2I3}, the results are pretty lengthy and can be found in supplementary notebook file.

\subsection{Analytic continuation}
The result \eqref{I2FinalResult} is correct for any $\text{Im}(p) \ge 0$ but it is not applicable at the point $p=0$ because $\kappa_{\pm}$ becomes infinite. In this section, we show how one can get an analytical answer for the case $p=0$. Straight away, there is a temptation to simply substitute $\kappa_{\pm}=\infty$ in the final formula, but this will lead to obviously wrong result. In order to find the correct expression, it is necessary to expand each eMPL in a Laurent series around the point $p=0$, however, this is difficult to do. The main complication is the presence of functions $Z_4(\kappa_{\pm},\vec{a})$ in the integration kernels that must be expanded along with the upper integration limit.  For this reason, we choose an easier way. Consider the sum of two kernels 
\begin{equation}
\Psi_{-1}(\kappa_+,x,\vec{a})+\Psi_{-1}(\kappa_-,x,\vec{a})=\frac{y(\kappa_+)}{y(x-\kappa_+)}+\frac{y(\kappa_-)}{y(x-\kappa_-)}+\frac{c4}{y}\left(Z_4(\kappa_+,\vec{a})+Z_4(\kappa_-,\vec{a})\right).
\end{equation}

Now it is necessary to expand this expression in a Taylor series around the point $p=0$ and take the corresponding limit, to do so we use the relation $y(\kappa_+)=y(\kappa_-)=\frac{m^2\eta}{p^2(s+m^2)}$ and the formula \eqref{z4ForI2}. We find 
\begin{equation}
\Psi_{-1}(\kappa_+,x,\vec{a})+\Psi_{-1}(\kappa_-,x,\vec{a})=2\Psi_{-1}(\infty,x,\vec{a})+4\pi i \Psi_{0}(0,x,\vec{a})+\mathcal{O}(p^2),
\end{equation}
which leads to the simple result
\begin{equation}
\Ef{... & -1 & ...}{... & \kappa_{+} & ...}{x}{a}+\Ef{... & -1 & ...}{... & \kappa_{-} & ...}{x}{a} = 2\Ef{... & -1 & ...}{... & \infty & ...}{x}{a}+4\pi i \Ef{... & 0 & ...}{... & 0 & ...}{x}{a} +\mathcal{O}(p^2).
\end{equation}
If we apply this relation to the answer \eqref{I2FinalResult} and compare with the answer \eqref{I1FinalResult}, we will immediately see the relation 
\begin{equation}
I_2^{(0)}\Big|_{s\rightarrow t}=I_1^{(0)}+\mathcal{O}(p^2).
\end{equation}
On the other hand, one can verify the correctness of the last equation simply by comparing the Symanzik polynomials \eqref{FforI1} and \eqref{FforI2} between themselves. Thus, we get one more confirmation of the correctness of our result.

\section{Diagram with five internal lines}
\label{I4}
\begin{figure}[h]
\center{\includegraphics[width=0.5\textwidth]{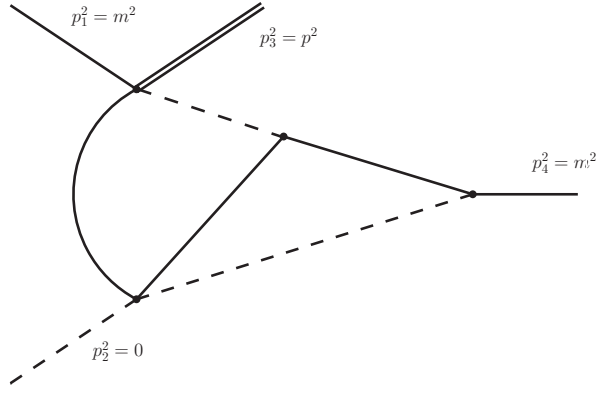}}
\caption{$I_4$ diagram. Dashed lines denote massless propagators and on-shell massless external particles; thick lines represent massive propagators and on-shell massive external particles; double line denotes off-shell external particle.}
\label{E11Diagram}
\end{figure}

In previous examples, we always integrated in the range $[0,\infty]$ except the last integration, where we used the M\"obius transformation $x \rightarrow \frac{x}{1-x}$. In order to substitute the integration limits, we used equations of the type \eqref{weight1Limit} and \eqref{weight2Limit}. However, it is possible to use a slightly different approach and perform the  M\"obius transformation for all integration variables, this can be convenient because integration limits becomes trivial. In this section, we show the application of this technique using the following Feynman integral as an example
\begin{equation}
I_4=\frac{e^{2\gamma_E\ep}(\mu^{2\ep})}{(i \pi^{d/2})^2}\int\frac{d^dk_1d^dk_2}{(k_1^2-m^2)((k_1-k_2)^2-m^2)((p_1-p_3+k_2)^2-m^2)(p_1-p_3+k_1)^2(p_2-k_2)^2}.
\end{equation}
The corresponding diagram can be found on the Figure \ref{E11Diagram}. 

The Symanzik polynomials are
\begin{equation}
U=x_2x_3+x_2x_4+x_1x_2+x_1x_3+x_1x_4+x_1x_5+x_2x_5+x_3x_5+x_4x_5,
\end{equation}
\begin{multline}
F=-tx_1\left(x_2x_3+x_2x_4+x_4x_3+x_4x_5\right)-\\
-m^2\left((x_1+x_2+x_3)(x_2x_3+x_1x_2+x_1x_3+x_2x_4+x_3x_4)+x_5(x_1x_4+(x_1+x_2)^2)\right),
\end{multline}
where $t=-(p_1-p_3)^2$, as before, we will work in the Euclidean region so that $t>0$.
And we can write the $\ep^0$ correction in terms of Feynman parametrization as 
\begin{equation}
I_4^{(0)}=\int\limits_0^{\infty}\frac{dx_1dx_2dx_3dx_4dx_5}{U F}\delta(1-x_2)=\int\limits_0^{\infty}\frac{dx_1dx_3dx_4dx_5}{U F |_{x_2=1}}.
\end{equation}

Performing the transformation $x_5=\frac{\bar{x}_5}{1-\bar{x}_5}$ and integrating over the variable $\bar{x}_5$ within the limits $[0,1]$ we obtain
\begin{equation}
I_4^{(0)}=\int\limits_0^{\infty}\frac{\left(-G\left(A_1;1\right)+G\left(A_2;1\right)\right)dx_1dx_3dx_4}{(x_4+x_3(1+x_1+x_4))(tx_1+m^2(x_3+x_1(1+x_3)+x_4+x_3x_4))},
\end{equation}
where $A_1$ and $A_2$ are the rational functions with respect to the variables $x_1$, $x_3$ and $x_4$:
\begin{equation}
A_1=\frac{x_3+x_1(1+x_3)+x_4+x_3x_4}{x_3(1+x_1+x_4)},
\end{equation}
\begin{equation}
A_2=\frac{tx_1(x_3+x_4+x_3x_4)+m^2(1+x_1+x_3)(x_1+x_3+x_4+x_1x_3+x_3x_4)}{tx_1x_3(1+x_4)+m^2\left(-1+x_3+x_1^2x_3+x_3^2+(1+x_3)^2x_4+x_1(x_3(3+x_3+x_4)-1)\right)}.
\end{equation}

After this we put  $x_4=\frac{\bar{x}_4}{1-\bar{x}_4}$ and bring all MPLs to the canonical form
\begin{equation}
I_4^{(0)}=\int\limits_0^{\infty}dx_1dx_3\int\limits_0^1d\bar{x}_4\frac{\left(-G\left(1+\frac{1}{x_1};\bar{x}_4\right)+G\left(B_1;\bar{x}_4\right)+G\left(B_2;\bar{x}_4\right)-G\left(B_3;1\right)-G\left(B_4;\bar{x}_4\right)+G\left(B_5;1\right)\right)}{\left(x_3(x_1(\bar{x}_4-1)-1)-\bar{x}_4\right)\left(tx_1(\bar{x}_4-1)+m^2(x_1(1+x_3)(\bar{x}_4-1)-x_3-\bar{x}_4)\right)},
\end{equation}
where $B_i, i=1...5$ are the rational functions with respect to the variables $x_1$ and $x_3$:
\begin{equation}
B_1=\frac{m^2(1+x_1)^2}{-tx_1+m^2(1+x_1+x_1^2)},
\end{equation}
\begin{equation}
B_2=\frac{x_1+x_3+x_1x_3}{-1+x_1+x_1x_3}, \qquad B_3=1+\frac{1+x_1}{-1+x_3+x_1x_3},
\end{equation}
\begin{equation}
B_4=-\frac{tx_1x_2+m^2(1+x_1+x_3)(x_1+x_3+x_1x_3)}{tx_1-m^2(1+x_1+x_3)(-1+x_1+x_1x_3)},
\end{equation}
\begin{equation}
B_5=-\frac{tx_1x_3+m^2(1+x_1+x_3)(x_1+x_3+x_1x_3)}{tx_1x_3+m^2(-1-x_1+x_3+x_1x_3(3+x_1)+(1+x_1)x_3^2)}.
\end{equation}

Further, in the same way, we integrate over the variable $x_3=\frac{\bar{x}_3}{1-\bar{x}_3}$. We do not give the appropriate calculations due to their large size. Having in mind that, if desired, they can be easily restored. And we arrive to the final integration
\begin{equation}
\label{I4FinalInt}
I_4^{(0)}=\frac{1}{m^2+t}\int\limits_0^{1}\frac{T_g+T_e}{x(x-1)}dx,
\end{equation}
where $T_g$ is the part completely free from the elliptic curve $y$
\begin{multline}
T_g=G\left(0,\xi\right) G(0,-\xi ,x)+G(\xi ,x) \left(G(\xi ,1)
   G\left(0,\xi\right)+G(0,\xi ,1)-\frac{\pi
   ^2}{3}\right)+G\left(0,\xi\right) G(0,\xi ,x)+
   \\
   +5
   G\left(0,\xi\right) G(\xi ,\xi ,x)+G\left(0,-\xi
   ,-\frac{m^2}{t},x\right)-2 G\left(\xi ,0,-\frac{m^2}{t},x\right)+G\left(\xi
   ,1,-\frac{m^2}{t},x\right)-
   \\
   -2 G(0,0,-\xi ,x)-2 G(0,0,\xi ,x)-2 G(0,-\xi ,0,x)-2
   G(0,\xi ,0,x)+G(0,\xi ,1,x)+
\\   
   +G(0,\xi ,\xi ,x)+G(\xi ,0,0,x)+2 G(\xi ,0,1,x)+G(\xi
   ,0,\xi ,x)-G(\xi ,1,0,x)-G(\xi ,1,1,x)+
\\   
   +6 G(\xi ,\xi ,0,x)-G(\xi ,\xi ,1,x),
\end{multline}
with $\xi=\frac{m^2}{m^2+t}$ and $T_e$ is the part which contain the elliptic curve
\begin{equation}
T_e=-R(0)+R\left(\frac{(m^2+t)x}{m^2(x-1)+tx}\right)+R\left(\frac{m^2}{x(m^2+t)}\right)-R\left(\frac{m^2(x+1)+tx}{(m^2+t)x}\right),
\end{equation}
where we introduce the shorten notations
\begin{equation}
R(a)=G(a,1,\Gamma(y);1)+G(a,1,\Gamma(-y);1)+G(a,\Gamma(y),1;1)+G(a,\Gamma(-y),1;1),
\end{equation}
and
\begin{equation}
\Gamma(y)=\frac{m^2\left(x^2-x-1-y\right)+t\left(x^2-x-y\right)}{m^2\left(x^2-3x+1-y\right)+t\left(x^2-x-y\right)}=\frac{\Upsilon(y)}{1+\Upsilon(y)}.
\end{equation}

The result of the final integration in \eqref{I4FinalInt} is very lengthy so we put the final answer for $I_4^{(0)}$ in supplementary notebook file.

\section{Next to linear reducible example} 
\label{I5}

\begin{figure}[h]
\center{\includegraphics[width=0.5\textwidth]{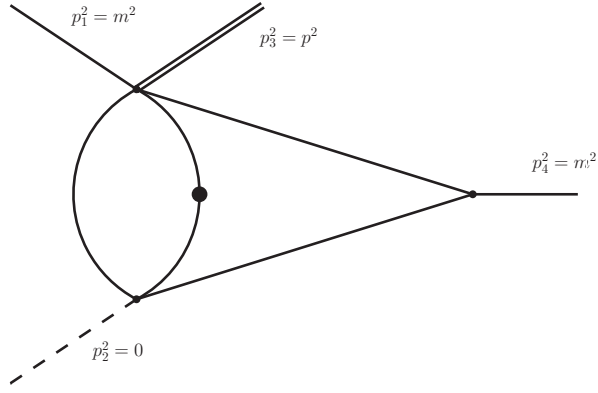}}
\caption{$I_m$ diagram. Dashed lines denote massless propagators and on-shell massless external particles; thick lines represent massive propagators and on-shell massive external particles; double line denotes off-shell external particle.}
\label{E3Diagram}
\end{figure}
As our last example, we will calculate the diagram shown on Figure \ref{E3Diagram}. It is easy to see that it is a generalization of the $I_1$ diagram from Figure \ref{E7Diagram}. The corresponding Feynman integral reads
\begin{equation}
I_m=e^{2\gamma_E\ep}(\mu^{2\ep})\int\frac{d^dk_1d^dk_2}{(i \pi^{d/2})^2}\frac{1}{(k_1^2-m^2)^2((k_1-k_2)^2-m^2)((p_1-p_3+k_2)^2-m^2)(p_2-k_2)^2-m^2}.
\end{equation}
The parametric representation is
\begin{equation}
I_m^{(0)}=\int\limits_0^{\infty}\frac{x_1 dx_1dx_2dx_3dx_4}{U F}\delta(1-x_1),
\end{equation}
where the Symanzik polynomials are
\begin{equation}
U=x_2x_3+x_2x_4+x_1x_2+x_1x_3+x_1x_4,
\end{equation}
\begin{equation}
F=-tx_1x_2x_3+m^2(x_1+x_2)x_3x_4-m^2\left(x_1+x_2+x_3+x_4\right)\left(x_2(x_3+x_4)+x_1(x_2+x_3+x_4)\right),
\end{equation}
with $t=(p_1-p_3)^2 > 0$.

Note that the second Symanzik polynomial $F$ is quadratic in all four parameters. In all previously discussed examples, Symanzik polynomials were quadratic in only three variables. In the work \cite{hidding2019all}, an integral with a similar situation was called next to linear reducible.

Integrating over $x_4$ we find:

\begin{multline}
\label{ImFurstInt}
I_m^{(0)}=-\int\limits_0^{\infty}\frac{dx_2dx_3}{y}\left(\frac{i \pi - G\left(-1;x_2\right)- G\left(0;q(y)\right)+G\left(0;x_2\right)+G\left(-1+\frac{1}{1+x_2};x_3\right)}{q(y)(1+x_2)+x_2+x_3+x_2x_3}\right.-
\\
\left.-\frac{i \pi - G\left(-1;x_2\right)- G\left(0;q(-y)\right)+G\left(0;x_2\right)+G\left(-1+\frac{1}{1+x_2};x_3\right)}{q(-y)(1+x_2)+x_2+x_3+x_2x_3}\right),
\end{multline}
where:
\begin{equation}
q(y)=\frac{-m(1+x_3+x_2(3+x_2+x_3))+y}{2(1+x_2)}
\end{equation}
and
\begin{equation}
\label{y2Polinomial}
y^2=-4 t x_2(1+x_2)x_3+m^2\left((1+x_2+x_2^2)^2-2(1+x_2)(1+3x_2+x_2^2)x_3-3(1+x_2)^2x_3^2\right).
\end{equation}

Next, we need to integrate over the variable $x_3$ or $x_2$, but we cannot do it right away because MPLs in the formula \eqref{ImFurstInt} contain arguments which are not rational in both $x_3$ and $x_2$ variables. The problem is the square root $y$ defined by \eqref{y2Polinomial}. Inside the root $y$ there is a polynomial in two variables.   
In order to perform the next integration, we need to find such rational coordinate transformation that rationalizes the $y$ root. Moreover, this must be done in two variables $x_3$ and $x_2$, otherwise on the next integration we may have an expression with nested radicals. In order to rationalize this root, we note that polynomial \eqref{y2Polinomial} is quadratic in the variable $x_3$ and have the form 
\begin{equation}
\label{rootRationalizable}
y^2=ax^2_3+bx_3+c^2,
\end{equation}
where $a$, $b$ and $c$ are polynomials in the variable $x_2$, it is particularly important that $c^2$ is a full square and $c=m(1+x_2+x_2^2)>0$.

It turns out that root \eqref{rootRationalizable} can be rationalized by a rational transformation, for this one can use the simple algorithm from work \cite{roots}. Note, that the algorithm from \cite{roots} cannot be applied directly to the curve \eqref{y2Polinomial} as to a curve in two variables $x_2$ and $x_3$ since, as a function of these two variables, it does not satisfy the algorithm conditions. Nevertheless, if we consider curve \eqref{y2Polinomial}  only as a function of the variable $x_3$, i.e. consider expression \eqref{rootRationalizable} then the algorithm is quite applicable. And we find simple relations 
\begin{equation}
x_3\rightarrow \frac{z(2c-bz)}{-1+az^2}, \qquad y\rightarrow c+\frac{2c-bz}{-1+az^2},
\end{equation}
where $z$ is a new variable.
Substituting the values for the $a$, $b$ and $c$ coefficients we get
\begin{equation}
x_3(z)\rightarrow-\frac{2 z\left(m(1+x_2+x_2^2)+2tx_2(1+x_2)z+m^2(1+x_2)(1+3x_2+x_2^2)z\right)}{1+3m^2(1+x_2)^2z^2},
\end{equation}
\begin{equation}
y\rightarrow\frac{-m(1+x_2+x_2^2)-2(1+x_2)(2t x_2+m^2(1+x_2(3+x_2)))z+3m^3(1+x_2)^2(1+x_2+x_2^2)z^2}{1+3m^2(1+x_2)^2z^2}.
\end{equation}
The Jacobian of this transformation
\begin{equation}
J_z=\frac{-2 m(1+x_2+x_2^2)-4(1+x_2)(2t x_2 +m^2(1+x_2(3+x_2)))z+6m^3(1+x_2)^2(1+x_2+x_2^2)z^2}{(1+3m^2(1+x_2)^2z^2)^2}.
\end{equation}
At the same time, for convenience, we will also perform the Mobius transformation
\begin{equation}
x_2(x)\rightarrow\frac{x}{1-x}, \qquad J_x=\frac{1}{(x-1)^2}.
\end{equation}
After this the limits of integrations becomes $x \in [0,1]$ and $z \in \left[0, \frac{i(x-1)}{\sqrt{3}m}\right]$.

Integrating over $z$ we find
\begin{multline}
\label{ImSecondInt}
I_m^{(0)}=\int\limits_0^{1}\frac{dx}{(m^2+t)x}\left\{\left(G(0;x)+G(1;x)\right)\left(G(f_1;\frac{i}{\sqrt{3}})-G(1;\frac{i}{\sqrt{3}})\right)+ G(0, \Theta(-y_2) ;\frac{i}{\sqrt{3}})+\right.
\\
 + G(0, \Theta(y_2) ;\frac{i}{\sqrt{3}})-G(0, \bar{\Omega}(-y_1) ;\frac{i}{\sqrt{3}})-G(0, \bar{\Omega}(y_1) ;\frac{i}{\sqrt{3}})-G(1, \Theta(-y_2) ;\frac{i}{\sqrt{3}})-G(1, \Theta(y_2) ;\frac{i}{\sqrt{3}})+
\\
+G(1, \Omega(-y_1) ;\frac{i}{\sqrt{3}})+G(1, \Omega(y_1) ;\frac{i}{\sqrt{3}})+G(f_1, \Theta(-y_2) ;\frac{i}{\sqrt{3}})+G(f_1, \Theta(y_2) ;\frac{i}{\sqrt{3}})-G(f_1, \Omega(-y_1) ;\frac{i}{\sqrt{3}})-
\\
\left.
-G(f_1, \Omega(y_1) ;\frac{i}{\sqrt{3}})-G(f_2, \Theta(-y_2) ;\frac{i}{\sqrt{3}})-G(f_2, \Theta(y_2) ;\frac{i}{\sqrt{3}})+G(f_2, \bar{\Omega}(-y_1) ;\frac{i}{\sqrt{3}})+G(f_2, \bar{\Omega}(y_1) ;\frac{i}{\sqrt{3}})
\right\},
\end{multline}
where:
\begin{equation}
\Theta(y_2)=\frac{m^2+m^2(x-1)x+2\sqrt{t}m y_2}{m^2(2-x+x^2)-4t(x-1)x},
\end{equation}
\begin{equation}
\Omega(y_1)=\frac{(t+m^2)((x-1)x+y_1)}{m^2(1-2x+x^2)-2t(x-1)x},
\end{equation}
\begin{equation}
\bar{\Omega}(y_1)=\frac{m^2(y_1-1)+t((x-1)x+y_1)}{m^2(2-x+x^2)+2t(x-1)x},
\end{equation}
and
\begin{equation}
f_1=\frac{m^2(1-x+x^2)}{m^2(1-(x-1)x)-2t(x-1)x},\qquad f_2=\frac{(t+m^2)(x-1)x}{m^2(2-x+x^2)-t(x-1)x}.
\end{equation}

The equation \eqref{ImSecondInt} contain two different roots of the fourth-degree polynomials. The first is the same root as discussed in section \ref{Section1} $y_1^2=(x-a_1)(x-a_2)(x-a_3)(x-a_4)$ with the brunch points \eqref{sunsetBP}. The properties of this root are well known to us. 

The second elliptic structure is
\begin{equation}
y_2^2=\frac{4t(x-1)^2x^2+m^2}{4t}=(x-b_1)(x-b_2)(x-b_3)(x-b_4),
\end{equation}
with the branch points 
\begin{equation}
\label{secondEC}
b_1=\frac{1}{2}(1+\varphi),\qquad b_2=\frac{1}{2}(1-\bar{\varphi}),\qquad b_3=\frac{1}{2}(1-\varphi),\qquad b_4=\frac{1}{2}(1+\bar{\varphi}),
\end{equation}
where we introduce the definitions $\varphi=\sqrt{1-\frac{2m}{\sqrt{-t}}}$, $\bar{\varphi}=\sqrt{1+\frac{2m}{\sqrt{-t}}}$.

In order to truly make sure that $y_1$ and $y_2$ are different elliptic curves, it is necessary to calculate the $j$ invariant using the equation \eqref{jInvariant}, the results are as follows
\begin{equation}
j_1=\frac{(3m^2+t)^3(3m^6+3m^4t+9m^2t^2+t^3)^3}{m^{12}t^2(m^2+t)^3(9m^2+t)}, \qquad j_2=\frac{256(3m^2+t)^3}{m^4(4m^2+t)}.
\end{equation}
So we see that $y_1$ and $y_2$ are really two different curves.

Result \eqref{ImSecondInt} has three features, the first is the presence of two elliptical structures. All Feynman integrals considered earlier included only one elliptic structure. These elliptic structures correspond to the two different sunsets in the form of subgraphs on Figure \ref{E3Diagram}. The second feature is that the elliptic curve corresponding to the sunset diagram now come in different combinations $\Omega(y_1)$, $\bar{\Omega}(y_1)$ and $\Theta(y_2)$. In all integrals discussed in previous chapters, the elliptic curve was included only in combination $\Upsilon(y)$. And the third important property is that the two elliptic structures $y_1$ and $y_2$ do not "mix" with each other. By this, we mean that the formula \eqref{ImSecondInt} does not contain such MPLs in which both $y_1$ and $y_2$ would be simultaneous. This third feature allows us to use the methods from\cite{broedelPure, Broedel:2019hyg} just as we did in all previous sections. 

Rewriting all MPLs through a linear combination of eMPLs, we find the following surprisingly simple answer
\begin{equation}
\label{IMAnswer}
I_m^{(0)}=\frac{1}{2(t+m^2)}\left[I_++I_-+I_+^*+I_-^*-4\pi i I_0 + 2I_G\right],
\end{equation}
were $I_{\pm}$, $I_{\pm}^*$ and $I_0$ are elliptic parts
\begin{equation}
I_{\pm}=\Ef{1 & -1 & -1}{0 & \alpha_{\pm} & 0}{1}{a}+\Ef{1 & -1 & -1}{0 & \alpha_{\pm} & 1}{1}{a}+2\Ef{1 & -1 & -1}{0 & \alpha_{\pm} & \infty}{1}{a},
\end{equation}
\begin{equation}
I_{\pm}^*=\Ef{1 & -1 & -1}{0 & \alpha_{\pm}^* & 0}{1}{a}+\Ef{1 & -1 & -1}{0 & \alpha_{\pm}^* & 1}{1}{a}+2\Ef{1 & -1 & -1}{0 & \alpha_{\pm}^* & \infty}{1}{a},
\end{equation}
\begin{equation}
I_{0}=\Ef{1 & 0 & -1}{0 & 0 & 0}{1}{a}+\Ef{1 & 0 & -1}{0 & 0 & 1}{1}{a}+2\Ef{1 & 0 & -1}{0 & 0 & \infty}{1}{a}
\end{equation}
and $I_G$ is the part which can be entirely written in terms of ordinary MPLs
\begin{multline}
I_G=\frac{1}{3} i \pi  G\left(0,\alpha _-,1\right)+\frac{1}{3} i \pi  G\left(0,\alpha
   _+,1\right)-\frac{1}{3} i \pi  G\left(0,\alpha
   _-^*,1\right)-\frac{1}{3} i \pi  G\left(0,\alpha
   _+^*,1\right)-
   \\
   -G\left(0,0,\alpha _-,1\right)-G\left(0,0,\alpha
   _+,1\right)-G\left(0,0,\alpha _-^*,1\right)-G\left(0,0,\alpha
   _+^*,1\right)+\frac{1}{2} G\left(0,\alpha _-,1,1\right)+
   \\
   +\frac{1}{2}
   G\left(0,\alpha _+,1,1\right)+\frac{1}{2} G\left(0,\alpha
   _-^*,1,1\right)+\frac{1}{2} G\left(0,\alpha _+^*,1,1\right)+2
   G\left(0,0,\sqrt[3]{-1},1\right)+
   \\
   +2
   G\left(0,0,-(-1)^{2/3},1\right)-G\left(0,\sqrt[3]{-1},1,1\right)-G\left(0,-(-1)^{2/3}
   ,1,1\right).
\end{multline}
In the higher equations, we introduced the notations
\begin{equation}
\alpha _{\pm}=\frac{1}{2}\pm\frac{1}{2} \sqrt{\frac{-m^4+3 m^2 t+2 i \sqrt{3} m^2
   \left(m^2+t\right)+t^2}{m^4+m^2 t+t^2}},
\end{equation}
\begin{equation}
\alpha _{\pm}^*=\frac{1}{2}\pm\frac{1}{2} \sqrt{\frac{-m^4+3 m^2 t-2 i \sqrt{3} m^2
   \left(m^2+t\right)+t^2}{m^4+m^2 t+t^2}}.
\end{equation}

Result \eqref{IMAnswer} is quite interesting since the second elliptical structure associated with $y_2$ in it is completely reduced. This feature is not obvious from the expression \eqref{ImSecondInt} since one can not cancel the dependence from $y_2$ in it. We cannot yet say whether this result is an accident or a consequence of some principle. 

\subsection{Discussions}  
\begin{figure}[h]
\center{\includegraphics[width=0.8\textwidth]{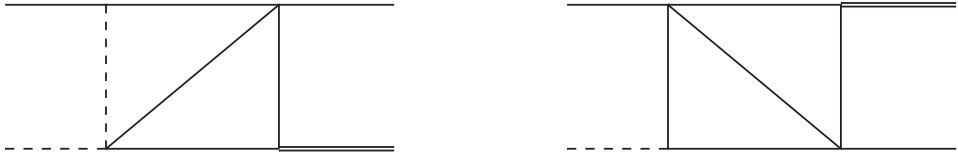}}
\caption{An example of diagrams for which the corresponding integrals will contain complex roots after the second(left) and first(right) integration.}
\label{eeDiagrams}
\end{figure} 
We managed to get the result \eqref{IMAnswer} for two main reasons. 

The first reason is that the curve \eqref{y2Polinomial} is quadratic in the variable $x_3$. This shape of the curve allowed us to find a rational transformation that rationalizes the root $y$.  But in many cases, we will face a more difficult situation. 
By way of example consider the diagram on the left part of Figure \ref{eeDiagrams}. In order to calculate it, it is necessary to take a nontrivial integral over four Feynman parameters (excluding one trivial). The first two integrations can be taken in terms of ordinary MPLs in a standard way, just as it was described in previous chapters. After that, we get an expression that contains the root of the polynomial in two variables. The situation is very similar to the case \eqref{y2Polinomial}, but now, this polynomial will be in the fourth power in both variables, which will not allow us to find its rationalizing transformation so easy. If we look at the diagram on the right side of Figure \ref{eeDiagrams}, the situation there is even worse. Already after first integration, there will appear a root of a fourth-degree polynomial in three variables. Therefore, we need to develop new methods for rationalizing the roots of complicated polynomials.

The second reason is that the two elliptic structures $y_1$ and $y_2$ in expression \eqref{ImSecondInt} do not mix with each other. This property is already violated when we consider the first $\ep$ correction $I_m^{(1)}$. Moreover, it is obvious that this will happen in higher orders of $\ep$ corrections. As a toy example consider simple function with mixing of two elliptic structures $G\left(\Theta(y_2);\Omega(y_1)\right)$. In order to rewrite it in terms of eMPLs, it must first be differentiated with respect to the variable x and then integrated back. The full derivative is 

\begin{equation}
\label{foolDer}
\frac{d G\left(\Theta(y_2);\Omega(y_1)\right)}{dx}=R(x,y_1,y_2)=R_1(x)+\frac{1}{y_1}R_2(x)+\frac{1}{y_2}R_3(x)+\frac{1}{y_1y_2}R_4(x), 
\end{equation}
where $R_i(x)$ denote the rational functions, the explicit expressions for those functions are not important for us, it's important that $R_4(x) \ne 0$. 

We see that formula \eqref{foolDer} contains the product of two roots $y_1y_2$ which is essentially a root of the eighth-degree polynomial. Integrals containing such roots are usually called hyperelliptic, for example, see \cite{byrd1954hyperelliptic}.  Thus, for this case, the eMPLs are no longer enough and we need a new class of functions which will be iterated integrals with hyperelliptic kernels. These functions should be introduced as a natural generalization of eMPLs and should contain them as a special case. The problem of introducing these functions and studying their properties will be the object of our future work. 

\section{Conclusions}
\label{Conclusions}
In this paper, we use methods from \cite{broedelPure,Broedel:2019hyg,broedelII} to analitycally calculate some two-loop Feynman integrals that are important when considering the processes of production and decay of heavy quarkonium\cite{chen2017two, chen2018two}. These Feynman integrals contain elliptic structures and are ultimately expressed through a set of pure eMPLs. All our results were verified numerically. In the last chapter, we pointed out two important problems arising in the calculation of complicated two-loop diagrams. The first problem is related to the rationalization of complex roots, and the second problem is the need to introduce a new class of functions. Both of these issues will be the subject of our future research.
\section*{Acknowledgements}
I would like to thank A.I. Onishchenko for interesting
and stimulating discussions as well as for general guidance in writing this work.
This work was supported by Foundation for the Advancement of Theoretical Physics and Mathematics "BASIS".

\section{Appendix A}
\label{AppendixA}
In this appendix we will prove the following statement: If all the branch points $a_i$ of the elliptic curve $y^2=(x-a_1)(x-a_2)(x-a_3)(x-a_4)$ are pairwise complex conjugates($a_1 = a_2^*, ~a_3 = a_4^* $) and two of the branch points satisfy the relation $1-a_1-a_3=0$, then for any $b \in \mathbb{C}$ we have the following relation
\begin{equation}
\label{theorem}
Z_4\left(\frac{1}{2}+b,\vec{a}\right)+Z_4\left(\frac{1}{2}-b,\vec{a}\right)=\frac{2 \pi i}{\omega_1},
\end{equation}
where $Z_4$ is defined in \eqref{Z4Eisenstein}.

To begin with, we prove that if the conditions of the theorem are satisfied, then $\mathit{z}_{*}$ defined in equation \eqref{Torus_coordinates} is expressed as follows
\begin{equation}
\label{zStarr}
\mathit{z}_{*}=\frac{c_4}{\omega_1}\int\limits_{a_1}^{-\infty}\frac{dx}{y}=\frac{1}{4}(-1+\tau),
\end{equation}
where $\tau=\frac{\omega_2}{\omega_1}$ is the module of the elliptic curve.
To prove this part we write $\mathit{z}_{*}$ as the direct sum of complex and imaginary parts $\mathit{z}_{*}=\frac{\mathit{z}_{*}+\mathit{z}_{*}^*}{2}+\frac{\mathit{z}_{*}-\mathit{z}_{*}^*}{2}$. Further, from the fact that the branch points $a_i$ are pairwise complex conjugates to each other we find that $ c_{4}$ , $\omega_1$ and the elliptic curve $y$ are real. Using this last part  it is easy to show, furthermore, that
\begin{equation}
\mathit{z}_{*}^*=\left(\frac{c_4}{\omega_1}\int\limits_{a_1}^{-\infty}\frac{dx}{y}\right)^*=\frac{c_4}{\omega_1}\int\limits_{a_2}^{-\infty}\frac{dx}{y}.
\end{equation}
First, lets calculate the real part $\text{Re}(\mathit{z}_{*})=\frac{\mathit{z}_{*}+\mathit{z}_{*}^*}{2}$ 
\begin{multline}
\frac{\mathit{z}_{*}+\mathit{z}_{*}^*}{2}=\frac{c_4}{2\omega_1}\int\limits_{a_1}^{-\infty}\frac{dx}{y}+\frac{c_4}{2\omega_1}\int\limits_{a_2}^{-\infty}\frac{dx}{y}=\frac{c_4}{2\omega_1}\int\limits_{a_1}^{-\infty}\frac{dx}{y}+\frac{c_4}{2\omega_1}\int\limits_{a_2}^{a_3}\frac{dx}{y}+\frac{c_4}{2\omega_1}\int\limits_{a_3}^{-\infty}\frac{dx}{y}=
\\
=\frac{1}{4}+\frac{c_4}{2\omega_1}\int\limits_{a_1}^{-\infty}\frac{dx}{y}-\frac{c_4}{2\omega_1}\int\limits_{a_1}^{\infty}\frac{dx}{y}=\frac{1}{4}-\frac{c_4}{2\omega_1}\int\limits_{-\infty}^{\infty}\frac{dx}{y}=-\frac{1}{4},
\end{multline}
where we have used the relation
\begin{equation}
\int\limits_{a_3}^{-\infty}\frac{dx}{y}=-\int\limits_{a_1}^{\infty}\frac{dx}{y}.
\end{equation}
The last one follows directly from the condition $1-a_1-a_3=0$. 

Similarly to the previous one, we find that $\frac{\mathit{z}_{*}-\mathit{z}_{*}^*}{2}=\frac{1}{4}\tau$. Putting the real and imaginary parts together we arrive at the formula \eqref{zStarr}.

To continue the proof, we need to use definition \eqref{Torus_coordinates} for $\mathit{z}_x$ and  establish a connection between $\mathit{z}_{\frac{1}{2}+ b}$ and $\mathit{z}_{\frac{1}{2}- b}$ functions. To do so we will use the Leibniz integral rule to find the full derivatives of $\mathit{z}_{\frac{1}{2}\pm b}$ with respect to $b$
\begin{equation}
\label{der_z}
\frac{d}{db}\mathit{z}_{\frac{1}{2}+b}=\frac{c_4}{\omega_1 y\left(\frac{1}{2}+b\right)}, \qquad
\frac{d}{db}\mathit{z}_{\frac{1}{2}-b}=-\frac{c_4}{\omega_1 y\left(\frac{1}{2}-b\right)}.
\end{equation}

Further, we note that if relation $1-a_1-a_3=0$ holds, then obviously relation $1-a_2-a_4=0$ also holds because $a_1=a_2^*, ~a_3=a_4^*$. Using this two equations it is easy to check that
\begin{equation}
\label{yrel}
y\left(\frac{1}{2}+b\right)=y\left(\frac{1}{2}-b\right).
\end{equation}
Adding equations \eqref{der_z} together and with the aid of equation \eqref{yrel} we see that
\begin{equation}
\label{sumOfAbelMaps}
\mathit{z}_{\frac{1}{2}+b}+\mathit{z}_{\frac{1}{2}-b}=c,
\end{equation}
where $c$ is independent from $b$. Thus, to find $c$ it will be enough for us to get an expression of $\mathit{z}_{\frac{1}{2}+b}+\mathit{z}_{\frac{1}{2}-b}$ for some specific value of $b$. As such specific value we take $b=\frac{1}{2}-a_1$; then $\frac{1}{2}+b=1-a_1=a_3$ and $\frac{1}{2}-b=a_1$. It's obvious from definition that $\mathit{z}_{a_1}=0$ therefore we only need to find the value $\mathit{z}_{a_3}$. Using elementary transformations, we obtain
\begin{equation}
c=\mathit{z}_{a_3}=\frac{c_4}{\omega_1}\int\limits_{a_1}^{a_3}\frac{dx}{y}=\frac{c_4}{\omega_1}\left(\int\limits_{a_1}^{a_2}\frac{dx}{y}+\int\limits_{a_2}^{a_3}\frac{dx}{y}\right)=\frac{1}{2}(1+\tau).
\end{equation}

The rest of the proof is now more or less straightforward. We need to use Definition \eqref{Z4Eisenstein} and find the relations between functions $g^{(1)}\left(\mathit{z}_{\frac{1}{2}-b}+\mathit{z}_{*},\tau\right)$, $g^{(1)}\left(\mathit{z}_{\frac{1}{2}-b}-\mathit{z}_{*},\tau\right)$, $g^{(1)}\left(\mathit{z}_{\frac{1}{2}+b}+\mathit{z}_{*},\tau\right)$ and $g^{(1)}\left(\mathit{z}_{\frac{1}{2}+b}-\mathit{z}_{*},\tau\right)$. First of all we use \eqref{zStarr} and \eqref{sumOfAbelMaps} to connect the arguments of this functions with each other 
\begin{equation}
\mathit{z}_{\frac{1}{2}+b}+\mathit{z}_{*}=-\left(\mathit{z}_{\frac{1}{2}-b}+\mathit{z}_{*}\right)+\tau,
\end{equation}
\begin{equation}
\mathit{z}_{\frac{1}{2}+b}-\mathit{z}_{*}=-\left(\mathit{z}_{\frac{1}{2}-b}-\mathit{z}_{*}\right)+1.
\end{equation}
And using \eqref{EisensteinKroneckerParity} and \eqref{EisensteinKroneckerTranslations} we find
\begin{equation}
g^{(1)}\left(-\left(\mathit{z}_{\frac{1}{2}-b}-\mathit{z}_{*}\right)+1,\tau\right)=-g^{(1)}\left(\mathit{z}_{\frac{1}{2}-b}-\mathit{z}_{*},\tau\right),
\end{equation}
\begin{equation}
g^{(1)}\left(-\left(\mathit{z}_{\frac{1}{2}-b}+\mathit{z}_{*}\right)+\tau,\tau\right)=-g^{(1)}\left(\mathit{z}_{\frac{1}{2}-b}+\mathit{z}_{*},\tau\right)-2\pi i.
\end{equation}
Inserting this in to \eqref{Z4Eisenstein} we immediately get \eqref{theorem}, this concludes the proof of our statement.

The proved statement was of great use in our work. Indeed, both elliptic curves that we examined \eqref{sunsetBP} and \eqref{secondEC} satisfy the conditions of the theorem. And all pars of arguments arising in $Z_4$, namely ($0$; $1$), $\kappa_{\pm}$, $\alpha_{\pm}$ and $\alpha_{\pm}^*$ can be represented in the form $\frac{1}{2}\pm b$.

\bibliographystyle{ieeetr}
\bibliography{litr}

\end{document}